\documentclass[12pt,a4,pdftex,achemso]{article}

\usepackage{color}
\usepackage{soul}
\usepackage{cite}
\usepackage{amsmath}
\usepackage[dvips]{graphics}
\usepackage{graphicx}
\usepackage{multicol}

\advance \topmargin by -\headheight
\advance \topmargin by -\headsep
\advance \topmargin by -1.0cm
\evensidemargin \oddsidemargin
\evensidemargin \oddsidemargin
\advance\hoffset by -3mm  
\advance\voffset by  8mm  
\newlength{\figwidth}
\newlength{\figheigth}
\newcommand{\onehalf}{\frac{1}{2}}

\newcommand{\be}{\begin{equation}}
\newcommand{\ee}{\end{equation}}
\newcommand{\bea}{\begin{eqnarray}}
\newcommand{\eea}{\end{eqnarray}}
\newcommand{\nn}{\nonumber}

\newcommand{\RR}{\mathbf{R}}
\newcommand{\rr}{\mathbf{r}}

\newcommand{\dr}{d\mathbf{r}}

\newcommand{\kk}{\mathbf{k}}

\newcommand{\rhorom}{{\rho\left({\mathbf r},\Omega\right)}}

\newcommand{\sigrom}{{\sigma \left({\mathbf r},\Omega\right)}}
\newcommand{\Om}{\mathbf{\Omega}}

\newcommand{\F}{{\cal F}}

\newcommand{\MU}{\boldsymbol{\mu}}

\newcommand{\PP}{\mathbf{P}}
\newcommand{\EE}{\mathbf{E}}

\begin{document}

\begin{center}
{\Large A Molecular Density Functional Theory of Water}

\vspace{5mm}
{\bf Guillaume Jeanmairet$^a$, Maximilien Levesque$^{a,b}$, Rodolphe Vuilleumier$^a$, and Daniel Borgis$^{a}$\footnote{Corresponding author: daniel.borgis@ens.fr}}

\vspace{2mm}

$^a${\em P\^ole de Physico-Chimie Th\'eorique, \'Ecole Normale Sup\'erieure, UMR 8640 CNRS-ENS-UPMC, 24, rue Lhomond, 75005 Paris, France} \\
$^b${\em Laboratoire PECSA, UMR 7195 CNRS-UPMC-ESPCI, 75005 Paris, France}

\end{center}

\vspace{5mm}

\begin{abstract}
Three dimensional implementations of liquid state theories offer an efficient alternative to computer simulations for the atomic-level description of aqueous solutions in complex environments. In this context, we present a (classical) molecular density functional theory (MDFT) of water that is derived from first principles and is based on two classical density fields, a scalar one, the particle density,  and a vectorial one, the multipolar polarization density. Its implementation requires as input the partial charge distribution of a water molecule and  three measurable bulk properties, namely the structure factor and the k-dependent longitudinal and transverse dielectric constants. It has  to be complemented by a solute-solvent  three-body term that reinforces tetrahedral order at short range. The approach is shown to provide the correct three-dimensional microscopic solvation profile around various molecular solutes,  possibly  possessing H-bonding sites, at a computer cost two-three orders of magnitude lower than with explicit simulations.
\end{abstract}

\newpage

\begin{center}
\resizebox{14cm}{!}{\includegraphics{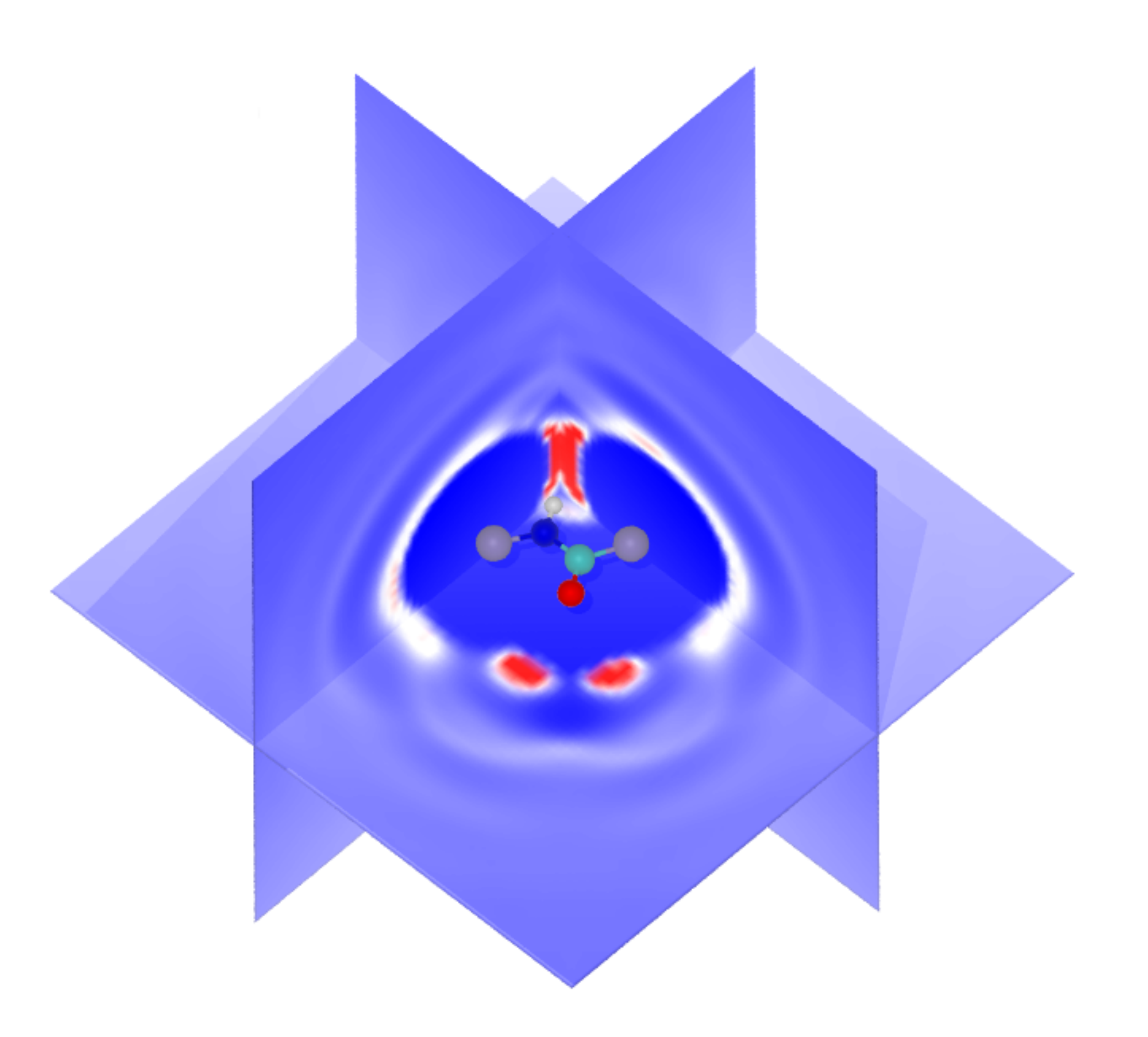}}

\noindent
TOC graphic
\end{center}


\newpage

\pagenumbering{arabic}

The numerical  methods that have emerged in the second part of the last century from  liquid-state theories\cite{hansen, gray-gubbins-vol1}, including integral equation theory in the interaction-site\cite{Chandler-RISM,hirata-rossky81,hirata-pettitt-rossky82,reddy03,pettitt07,pettitt08}
or molecular\cite{blum72a,blum72b,patey77,fries-patey85,richardi98,richardi99} picture, classical density functional theory (DFT)\cite{evans79,evans92,Wu07}, or classical fields theory\cite{chandler93,lum99,tenwolde01,coalson96}, have become methods of choice for 
many physical chemistry or chemical engineering applications\cite{gray-gubbins-vol2,neimark06,neimark11,wu06}. They can yield reliable predictions for both the microscopic structure and the thermodynamic properties of molecular fluids in bulk, interfacial, or confined conditions at a much more modest computational cost than  molecular-dynamics or Monte-Carlo simulations. 
A current challenge concerns their implementation in three dimensions in order to describe molecular liquids, solutions, and mixtures in  complex environments such as atomistically-resolved solid interfaces or biomolecular media. There have been a number of recent efforts in that direction using 3D-RISM\cite{Beglov-Roux97,kovalenko-hirata98,red-book,yoshida09,kloss08-jcp,kloss08-jpcb}, lattice field\cite{azuara06,azuara08} or Gaussian field\cite{tenwolde01,varilly11} theories.
Recently, a molecular density functional theory (MDFT) approach to solvation has been
introduced.~\cite{ramirez02,ramirez05-CP,ramirez05,gendre09,zhao11,borgis12} It relies on the definition of a free-energy functional depending on the full six-dimensional position and orientation solvent density. In the so-called homogeneous reference fluid (HRF) approximation, the (unknown) excess free energy can be inferred from the angular-dependent direct correlation function of the bulk solvent,
that can be predetermined from molecular simulations of the pure solvent. Compared to reference molecular dynamics calculations, such approximation was shown to be accurate for polar, non-hydrogen bonded fluids \cite{ramirez02,gendre09,zhao11,borgis12,levesque12_2}, but to require some corrections for water\cite{zhao11,zhao-wu11,zhao-wu11-correction,levesque12_1}.

In this paper we introduce a simplified version of MDFT that can be derived rigorously for SPC- or TIPnP-like representations of water, involving a  single Lennard-Jones interaction site and distributed partial charges. In that case we will seek a simpler functional form expressed in terms of the particle density $n(\rr)$ and site-distributed polarisation density $\PP(\rr)$, and requiring as input simpler physical quantities than the full position and  orientation-dependent direct correlation function. This functional may be considered as a multipolar generalization of the generic dipolar fluid free-energy functional $\F[n(\rr),\PP(\rr)]$ introduced in Refs.~\cite{ramirez02,ramirez05}.

\textit{A functional of $n(\rr)$ and $\PP(\rr)$.} 
We start from SPC- or TIPnP-like representation of water, constituted by a single Lennard-Jones center, located on
the oxygen, and $m$ charges distributed on various sites. Each molecule is supposed to be rigid with position $\rr_i$ and
orientation $\Omega_i$. For a given water configuration, we define the microscopic particle densities 
\bea
\label{eq:hatrhon}
  \hat{\rho}(\rr,\Omega) &= & \sum^N_{i = 1} \delta(\rr-\rr_i) \delta(\Omega - \Omega_i), \\
  \hat{\rho}_n (\rr) & =& \sum^N_{i = 1} \delta(\rr-\rr_i)  = \int d\Omega  \hat{\rho}(\rr,\Omega),
\eea
and the charge and multipolar polarization density 
\bea
\label{eq:hatrhoc}
  \widehat{\rho}_c (\rr) & = & \sum^N_{i = 1} \sigma(\rr -\rr_i, \Omega_i) \nn  \\
    & = & \int d\rr' d\Omega \, \sigma(\rr-\rr', \Omega) \hat{\rho}(\rr', \Omega), \\
    \widehat{\PP}_c (\rr) & = & \sum^N_{i = 1} \MU(\rr -\rr_i, \Omega_i)  \nn  \\
    & = & \int d\rr' d\Omega \, \MU(\rr-\rr', \Omega) \hat{\rho}(\rr', \Omega).
\eea
$\sigma (\rr, \Omega)$ is  the molecular charge density and,  according to the definition of  Refs~\cite{raineri93,bopp98}, $\MU(\rr, \Omega)$ is the molecular polarization density of a water molecule taken at the origin with orientation $\Omega$: 
\bea
\sigrom & = & \sum_m q_m \, \delta\left(  \rr - \mathbf{s}_m(\Omega)      \right),
\label{eq:sigma_water} \\
\MU(\rr,\Omega) &= & \sum_m q_m \, \mathbf{s}_m(\Omega)   \int_0^1 du \, \delta ( \rr - u \, \mathbf{s}_m(\Omega) ),
\eea
where  $\mathbf{s}_m(\Omega)$ indicates the location of the $m^{th}$ atomic site for a given $\Omega$.   It can be easily checked that molecular charge and polarization densities are linked by the usual relation $\sigma(\rr,\Omega) = -\mathbf{\nabla} \cdot \MU(\rr,\Omega) $. In k-space
\bea
\MU(\kk,\Omega) &= & -i \, \sum_m q_m \frac{\mathbf{s}_m(\Omega) }{\kk \cdot \mathbf{s}_m(\Omega) } \left( e^{i \, \kk \cdot \mathbf{s}_m(\Omega) } -1 \right) \\
\label{eq:mukom}
&= & \MU(\Omega) + \frac{i}{2} \sum_m q_m \left( \kk \cdot \mathbf{s}_m(\Omega) \right)  \mathbf{s}_m(\Omega) + ... ,
\eea
with $\MU(\Omega) = \sum_m q_m \, \mathbf{s}_j(\Omega) =  \mu_0 \, \Om$  ($\mu_0$ the molecule dipole; for water, $\Om$ is
the unit vector along the O-H bonds angle bissector), such that the molecular polarization density reduces to a molecular dipole located at the origin at dominant order, but does include  the complete multipole series at all other orders. 

Our derivation starts from the observation that the Hamiltonian of  $N$ water molecules in the presence of an embedded solute, described by an external  molecular force field,  can be written as
 \be
\label{eq:H_rho_rhoc}
H_N = T + U + \int \dr \, \hat{\rho}_n (\rr) \Phi_n(\rr) - \int \dr \, \hat{\PP}_c (\rr) \cdot \EE_c(\rr),
\ee
where $T$ and $U$ are  the water kinetic and  pair-wise potential energy, respectively. $\Phi_n(\rr), \, \EE_c(\rr)$ denote the value of the
external Lennard-Jones potential and electric field at position $\mathbf{r}$
\begin{eqnarray}
\label{eq:Phin}
  \Phi_n (\mathbf{r}) & = & \sum_j 4 \varepsilon_j \left[ \left(
  \frac{\sigma_j}{| \mathbf{r}-\mathbf{R}_j |}  \right)^{12} - \left(
  \frac{\sigma_j}{| \mathbf{r}-\mathbf{R}_j |}  \right)^6 \right],\\
  \label{eq:Ec}
  \EE_c (\mathbf{r}) & = & \frac{1}{4 \pi \epsilon_0} \sum_j Q_j \, \frac{\mathbf{r}-\mathbf{R}_j}{| \mathbf{r}-\mathbf{R}_j |^3}.
\end{eqnarray}
The  solute is here described by atomic sites $j$, located at $\mathbf{R}_j$, with Lennard-Jones
parameters $\sigma_j, \varepsilon_j$ (using Lorentz-Berthelot mixing rules with respect to water), and point
charges $Q_j$. 

Following  the original derivation of Evans\cite{evans79} and extending  it to four independent external field variables, 
$\Phi_n(\rr)$ and $E_{x,y,z}(\rr)$, instead of just one, it can be easily proved  that the grand potential $\Theta$ of the solute-water system
at a given water chemical potential may be expressed as a functional of the one-particle number density, $n(\rr) =  \left< \rho_n(\rr) \right>$,  and of the one-particle polarization density, $\PP(\rr) =  \left< \hat{\PP}_c (\rr) \right>$, that is $\Theta = \Theta[n(\rr),\PP(\rr)]$. Minimization of this functional with respect to those two fields yields the equilibrium densities and the value of the grand potential.  Taking as a reference the bulk water system at the same chemical potential, with number density $n_0$ and grand potential $\Theta_0$, the same properties hold true for the solvation free energy 
$\F[n(\rr),\PP(\rr)] = \Theta[n(\rr),\PP(\rr)] - \Theta_0$.  This constitutes the first important result of this work.

\begin{figure}
\begin{center}
\resizebox{16cm}{!}{\includegraphics{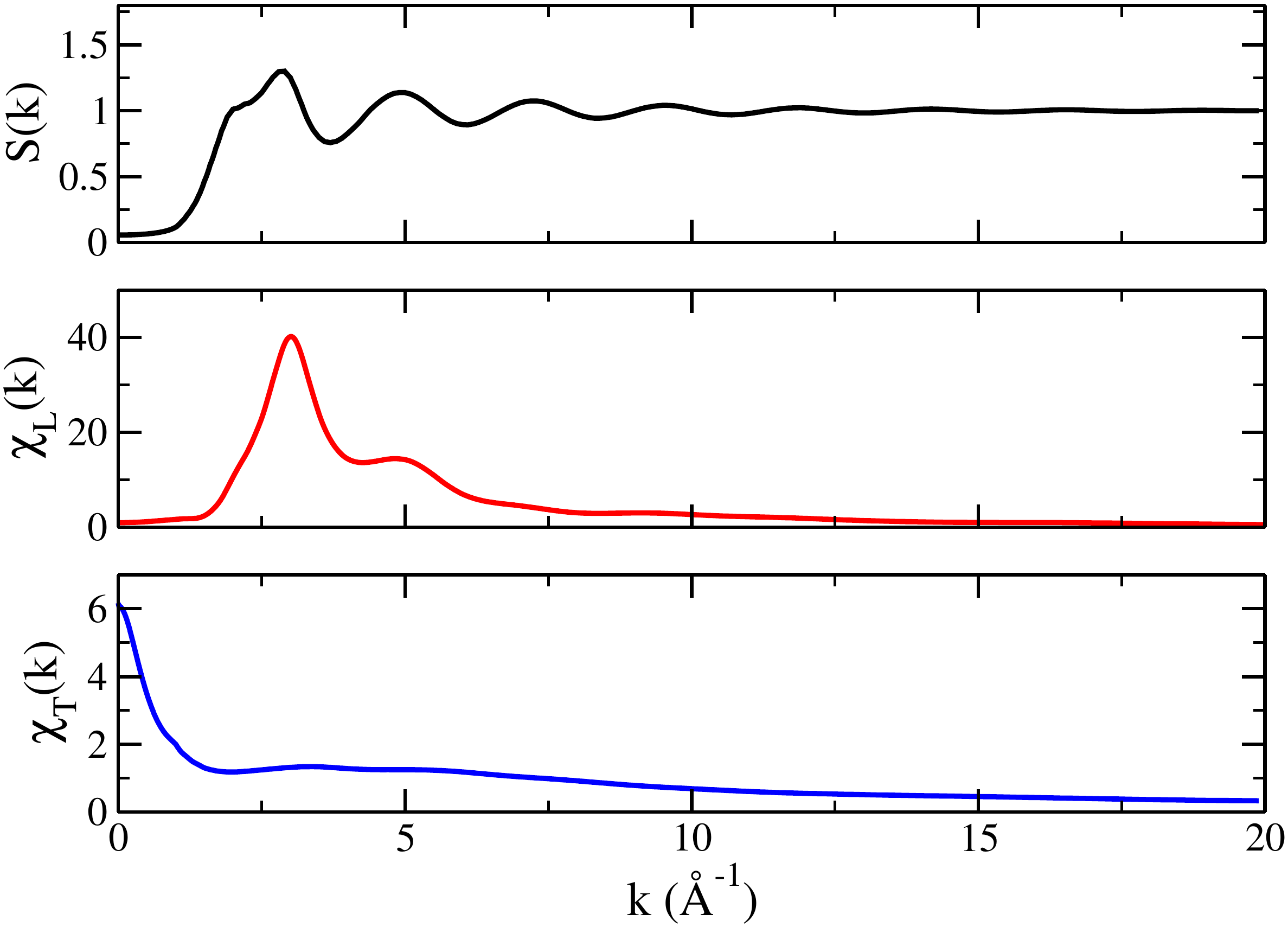}}
\end{center}
\caption{
\label{fig:S_and_chi_SPCE}
Structure factor and longitudinal and transverse dielectric susceptibility for SPC/E water computed according to the
formulas in Ref~\cite{bopp98}
}
\end{figure}

\textit{Expression of the functional.}
We proceed by invoking an equivalent of the Kohn-Sham decomposition in electronic DFT. Introducing the intrinsic Helmholtz 
free energy
\be
\label{eq:Ftot}
\F[n(\rr),\PP(\rr)] = \F_{int}[n(\rr),\PP(\rr)] + \int \dr \, \Phi_n(\rr) n(\rr) - \int \dr \, \EE_c(\rr) \cdot \PP(\rr) 
\ee
and decomposing it into  ideal (non-interacting) part and excess part, $\F_{int} = \F_{id} +  \F_{exc}$, we go back to the full one-particle angular density $\rhorom = \left< \hat{\rho}(\rr,\Omega) \right>$ to express the ideal term (equivalent to going back to orbitals in eDFT)
\be 
\label{eq:Fid}
\F_{id}[\rho]  =
k_BT \int d\rr d\Omega \left[ \rhorom \ln \left(\frac{8\pi^2 \rhorom}{n_0} \right) - \rhorom)+\frac{n_0}{8\pi^2} \right], 
\ee
where $d\Omega$ indicates an angular integration over the three Euler angles. The spatially and orientationally uniform fluid density $n_0/8 \pi^2$ is used as reference. We also expand the excess term around the homogeneous liquid state at that density
\bea
\F_{exc}[n(\rr),\PP(\rr)] & =  & \F_{int}[n(\rr),\PP(\rr)] - \F_{id}[n(\rr),\PP(\rr)] \nn \\
& = & \frac{k_B T}{2 }\int d\rr_1 d\rr_2 \,  S^{-1}(r_{12}) \, \Delta n(\rr_1) \, \Delta n(\rr_2)  \nn \\
& + & \frac{1}{8 \pi \epsilon_0
} \int d\rr_1 d\rr_2  \, \chi_L^{-1}(r_{12}) \, \PP_L(\rr_1) \cdot  \PP_L(\rr_2)  
\label{eq:Fexc} \\
& + & \frac{1}{8 \pi \epsilon_0} \int d\rr_1 d\rr_2  \, \chi_T^{-1}(r_{12}) \, \PP_T(\rr_1) \cdot  \PP_T(\rr_2)\nn \\
& - & k_B T \int d\rr \frac{\Delta n(\rr)^2}{2n_0} - k_B T \int d\rr \, \frac{3}{2 \mu_0^2 n_0} \PP(\rr)^2 + \F_{cor}[n(\rr),\PP(\rr)]. \nn 
\eea
The first three terms stem from the quadratic expansion of $\F_{int}$. Here $\PP_L, \PP_T$ designate the longitudinal anrd transverse components of the polarization vector, that are defined in k-space by
\be
\PP_L(\kk) = (\PP(\kk) \cdot \hat{\kk})  \hat{\kk}, \,  \, \PP_T(\kk) = \PP(\kk) - \PP_L(\kk).
\label{eq:P_L_T}
\ee
$S^{-1}(r)$ and $\chi_{L,T}^{-1}(r)$ are defined as the Fourier transforms of the inverse of the structure factor $S(k)$ and longitudinal and transverse electric susceptibility $\chi_{L}(k)$ and $\chi_{T}(k)$, which may be obtained either from experimental data or by simulations through the correlation of the molecular density fluctuations and of the longitudinal and transverse polarisation vector fluctuations\cite{raineri92,raineri93,bopp96,bopp98}. We have computed these quantities for SPC/E water \cite{berendsen87} using the k-space method described by Bopp et al\cite{bopp96,bopp98}; see Fig.~\ref{fig:S_and_chi_SPCE}. The dielectric susceptibilities are related to the longitudinal and transverse dielectric constants through $\chi_L(k) = 1 - 1/\epsilon_L(k)$ and 
$\chi_T(k) = (\epsilon_T(k) - 1)/4\pi$.   Their very small k behavior, impaired in simulation by finite size effects,  can thus be extrapolated using the macroscopic value of the dielectric constants $\epsilon_L(0)
= \epsilon_T(0) = 71$\cite{kusalik94}. 

Note that in eqn~\ref{eq:Fexc} the longitudinal polarization term can be also written as a charge-charge interaction term involving the bound charge density $\rho_c(\rr) = - \nabla \cdot \PP(\rr)$ (that would also include the free charges in the case of a ionic solution).  Note also that we have neglected the density-polarisation correlations which we  verified to be very small in water.
The negative terms in eqn~\ref{eq:Fexc} stand for the linearized translational and orientational entropy, whereas the last term is the unknown correction term that contains supposedly all orders in $\Delta n(\rr)$ and $\PP(\rr)$ higher that 2. 

As will be seen below, the functional with 
$\F_{cor}= 0$  tends to underestimate  the tetrahedral order in the vicinity of the solute atomic sites giving rise to H-bonds with the solvent. Such defect was also present in our previous MDFT functional, involving the full angular-dependent c-functions in the excess functional\cite{zhao11}. To correct it,  we introduce as before a short-range three-body potential term borrowed from the coarse-grained water model of Molinero {\em et al.} for water and ions in water\cite{molinero09,molinero_ions09} that enforces some tetrahedral order around all atomic sites $k$ susceptible to give a H-bond interaction
\bea
\F_{cor}[n(\rr)] &=  & \onehalf k_B T \sum_k [  \lambda_k^{1}  \int d\rr_2 d\rr_3 \, f_k(r_{k2}) \, f_k(r_{k3})(\frac{\rr_{k2} \cdot \rr_{k3}}{r_{k2}r_{k3}}  - \cos \theta_0 )^2 \, n(\rr_2) \, n(\rr_3) \nn \\
& + & \lambda_k^{2} \int d\rr_2 d\rr_3 f_k(r_{k2}) \, f_w(r_{23})(\frac{\rr_{k2} \cdot \rr_{23}}{r_{k2}r_{23}}  - \cos \theta_0 )^2 \, n(\rr_2) 
\, n(\rr_3)  ],
\label{eq:F3b}
\eea
with $r_{ki} = |\rr_i - \RR_k|$.
The two terms describe three-body H-bonding to the the first and second water solvation shell, respectively. $f_k(r)$ and $f_w(r)$ are the short-range functions introduced by DeMille and Molinero for ion-water and water-water contributions. Their systematic parametrization, as well as the choice of the strength parameters $\lambda_k^1, \lambda_k^2$, were alluded to in Ref.~\cite{zhao11}. For simplicity,
we  have selected  below the values $k_BT \lambda_k^1 = k_BT \lambda_k^2 = 75$ kJ/mol for ionized H-bonding sites, and 50 kJ/mol for neutral ones, whatever their chemical nature. 

We note finally that the above functional \ref{eq:Ftot}-\ref{eq:Fexc} (with $\F_{cor}= 0$) can be shown to be equivalent  to the generic dipolar functional introduced  in Refs~\cite{ramirez02,ramirez05} if the molecular polarization density is restricted to dipolar order in eqn~\ref{eq:mukom}, such that $\PP(\rr) = \mu_0 \int d\Omega \, \Om \, \rhorom$. The present work can thus be viewed as  a generalization of Refs~\cite{ramirez02,ramirez05} to a Stockmayer-like multipolar solvent.

\textit{Implementation and results.}
The functional defined by eqs~\ref{eq:Ftot} to \ref{eq:Fexc}   can be minimized in the presence of the external potential by discretizing $\rhorom$ on a  position and orientation grid.  The minimization is performed in fact with respect to a ''fictitious wave function'', $\Psi(\rr,\Omega)=\rho(\rr,\Omega)^{1/2}$,  in order to prevent the density from becoming negative in the logarithm. The position are represented on a 3D-grid with typically 3-4  points per Angstrom whereas the orientations are discretized using a Lebedev grid (yielding the most efficient angular quadrature for the unit sphere\cite{lebedev99}) for $\theta, \phi$, the orientation of the molecule $C2v$ axis, and a regular grid for the remaining angle $\psi$ (from $0$ to $\pi$). We use typically $N_\Omega = 12 \times 3$ angles. The calculation begins with the tabulation of the external Lennard-Jones potential $\phi_n(\rr)$, and external electric field $\EE_c(\rr)$. The latter is computed by extrapolating the solute charges on the grid and solving the Poisson equation. The molecular dipole function $\MU(\kk,\Omega)$ is also precomputed in k-space.   At each minimization cycle, $\rhorom$ is Fourier transformed to get 
$\rho(\kk,\Omega)$,  which is integrated over angles to get $n(\kk)~=~\int d\Omega \, \rho(\kk,\Omega)$ and $\PP(\kk) = \int d\Omega \, \MU(\kk,\Omega) \rho(\kk,\Omega)$,  as well as  its longitudinal and transverse components defined by eqn~\ref{eq:P_L_T}.
The non-local excess terms in \ref{eq:Fexc} are computed in k-space and so are the gradients with respect to $\rho(\kk,\Omega)$; those are then back-transformed to r-space. The three-body contribution can be  expressed as a sum of one-body terms and computed efficiently  in r-space. For minimization, we used the L-BFGS quasi-Newton optimization routine\cite{BFGS} which requires as input at each cycle free energy and gradients. 
The calculation are performed on   a standard  workstation or laptop. The convergence turns out to be fast, requiring at most 25-30 iterations, so that despite the important number of FFT's be handled at each step ($2 \times N_{\Omega}$)\cite{fftw3}, one complete minimization takes only a few minutes on a single processor. The computation of identical quantities using  molecular dynamics simulations requires $10^2-10^3$ times more computer time\cite{borgis12,levesque12_2}.

To describe water, we have chosen the SPC/E model\cite{berendsen87} which defines both the external potential in eqs~\ref{eq:Phin}-\ref{eq:Ec} and the structure factor and dielectric susceptibilities of equation~\ref{eq:Fexc}. Functional minimization was performed in the presence of different solutes placed the center of cubic box of size $L = 25 \, \AA$, with $80^3$ grid points.  In figure 2, we display the ion-oxygen pair distribution functions obtained by MDFT minimization around two representative monovalent ions and compare them to the corresponding curves generated by molecular dynamics simulations using a similar box size.  In the straight homogeneous reference fluid approximation, $F_{corr}=0$ in eqn~\ref{eq:Fexc}, it appears as in Ref.~\cite{zhao11} that the first peak exhibits the correct position and width but is overestimated in height, whereas the second peak is misplaced; this later deficiency points  to a lack of tetrahedral order. Those features are well corrected by inclusion of the three-body term of eqn~\ref{eq:F3b}.   Identical comparisons are carried out in figure 3 for a SPC/E water molecule dissolved in our MDFT water model, and in figure 4 for a N-methyl-acetamide molecule (NMA: CH$_3$NHCOCH$_3$), the paradigm for a peptide bond chemical motif. Three-body contributions were added for the O  and N sites with the parameters given in the previous section,  the H-atoms being  included implicitly in the associated heavy atom. The MDFT approach is able to  provide not only pair distribution functions, but also the full three-dimensional solvent structure; this is illustrated for each molecule by an image of the surrounding three-dimensional water density. For water in water, the expected tetrahedral,  four-fold coordination appears clearly, as does the fine structuring of water around the C=O and N-H bonds of  NMA.

The purpose of this note was to derive  a refined version of MDFT that is  based on two natural physical fields, the particle density and polarization density, and on three simple, measurable bulk water properties. It was shown also that an efficient three-dimensional numerical implementation can be carried out. The next step will be the application of this new formalism to the hydration of complex molecular solutes, such as the atomistically-resolved clay surfaces considered in Ref.~\cite{levesque12_2} with a reduced dipolar water model, or to the hydration of biological molecules. We will focus also on a systematic study of the  performance of our novel MDFT scheme  for the accurate  prediction of molecules hydration free energies, in addition to the microscopic hydration structure.

\noindent
{\bf \large Acknowledgments}\\
The authors acknowledge financial support from the Agence Nationale
de la Recherche under grant ANR-09-SYSC-012. DB acknowledges many thorough discussions about DFT with Rosa Ramirez and Shuangliang Zhao.


\newpage

\newpage

\begin{figure}
\begin{center}
\resizebox{16cm}{!}{\includegraphics{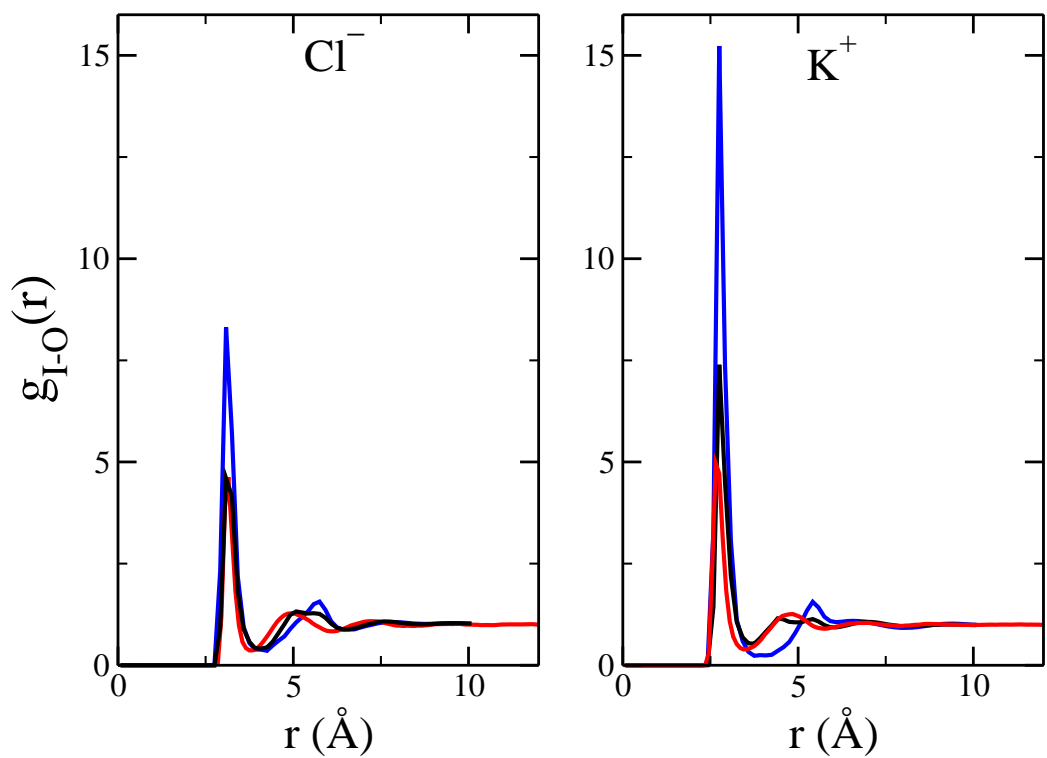}}
\end{center}
\caption{
\label{fig:g_Cl_Br}
 Ion-oxygen pair distribution functions  for chloride and potassium ions in SPC/E water computed by MD (red lines) or MDFT without (blue lines) or with the three body term of eqn~\ref{eq:F3b} (black lines)  }
\end{figure}

\newpage

\begin{figure}
\begin{center}
\hspace{2cm}\resizebox{8cm}{!}{\includegraphics{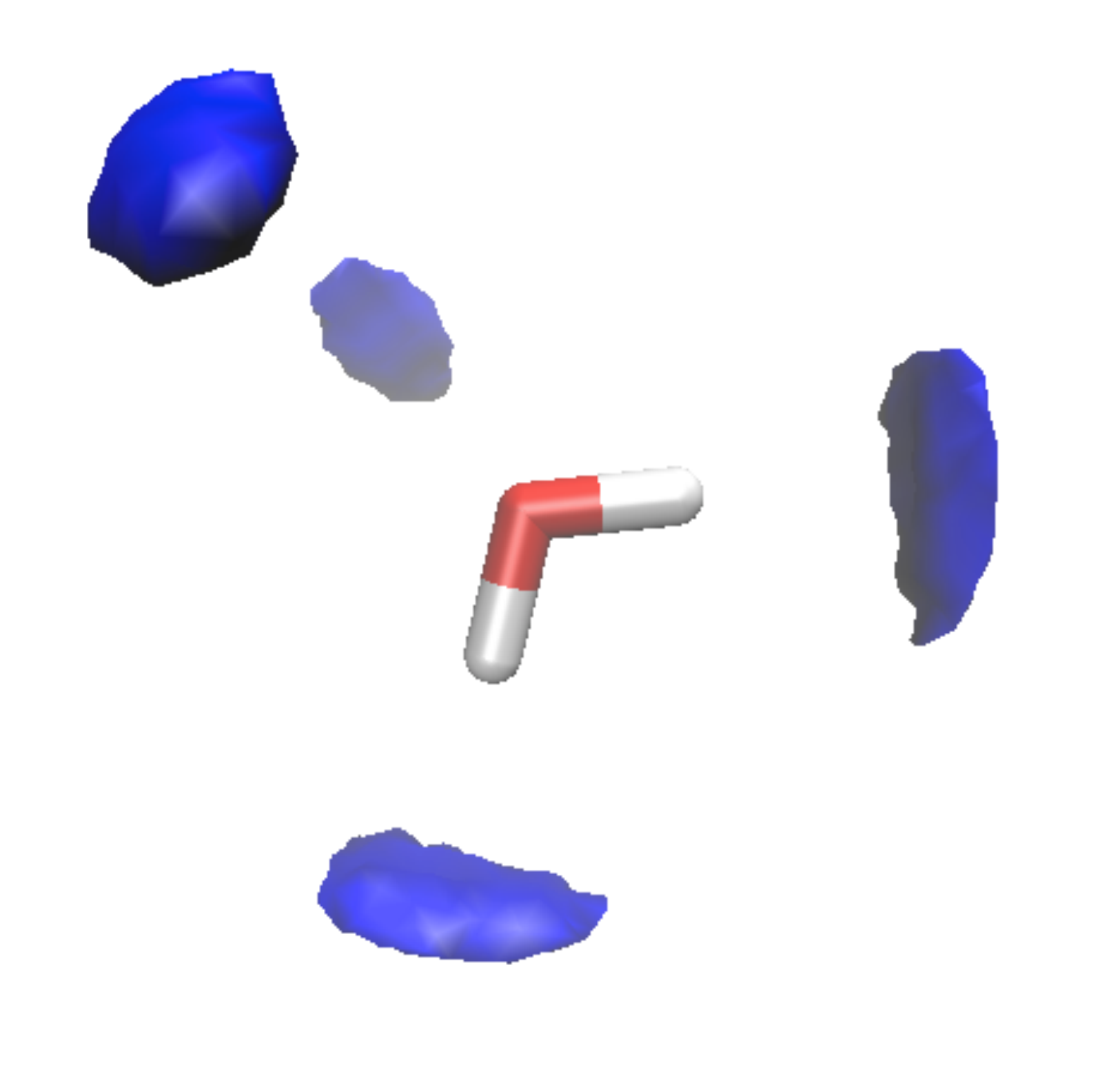}}
\vskip2em
\resizebox{14cm}{!}{\includegraphics{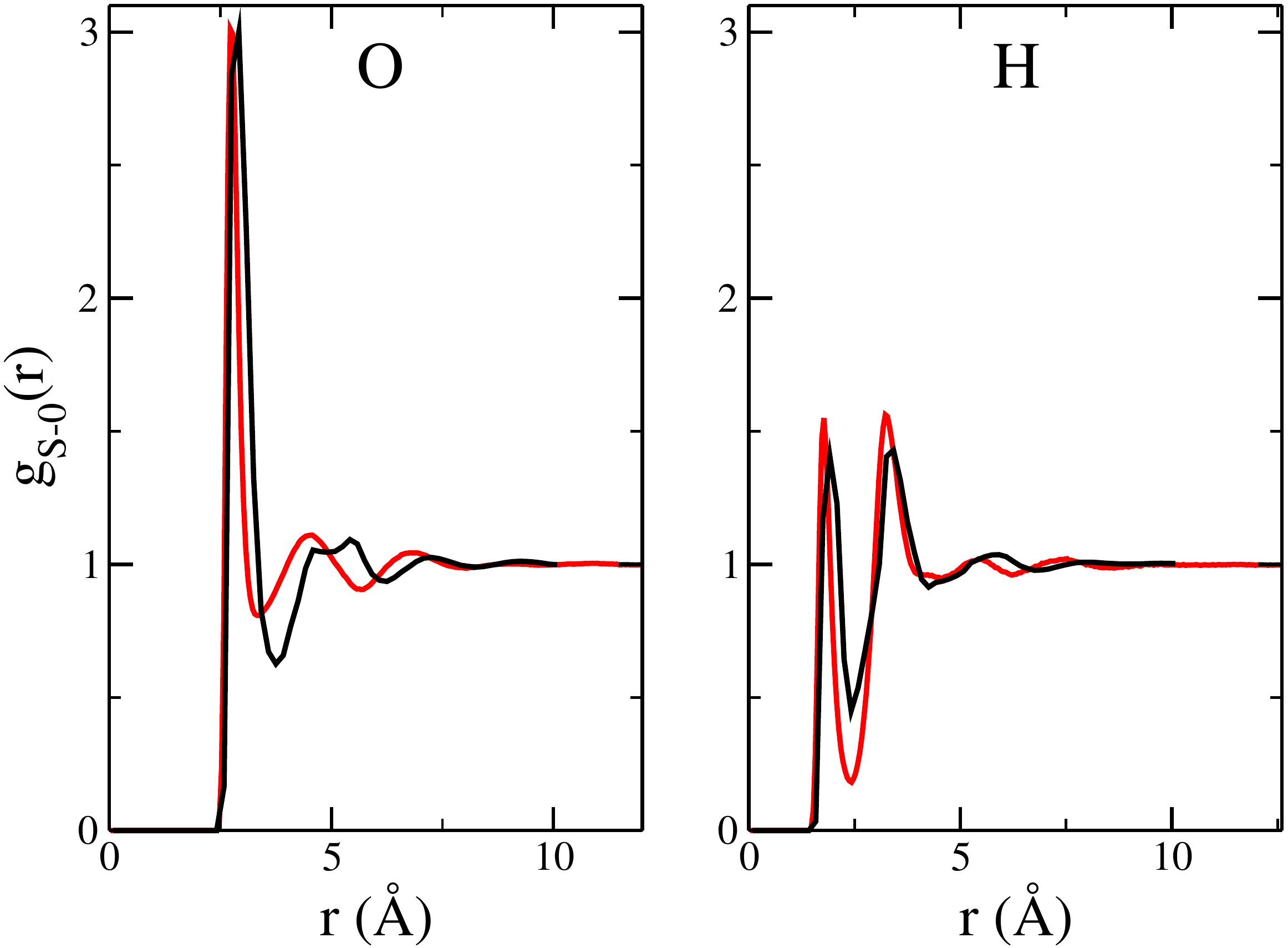}}
\end{center}
\caption{
\label{fig:gr_water}
Top: Isosurface of high water density ($n(\rr) = 3 n_0$) around a tagged SPC/E water molecule as determined by MDFT. Bottom: Corresponding site-oxygen pair distribution functions  computed by MD (red lines) or MDFT (black lines). }
\end{figure}

\newpage

\begin{figure}
\begin{center}
\hspace{2cm}\resizebox{10cm}{!}{\includegraphics{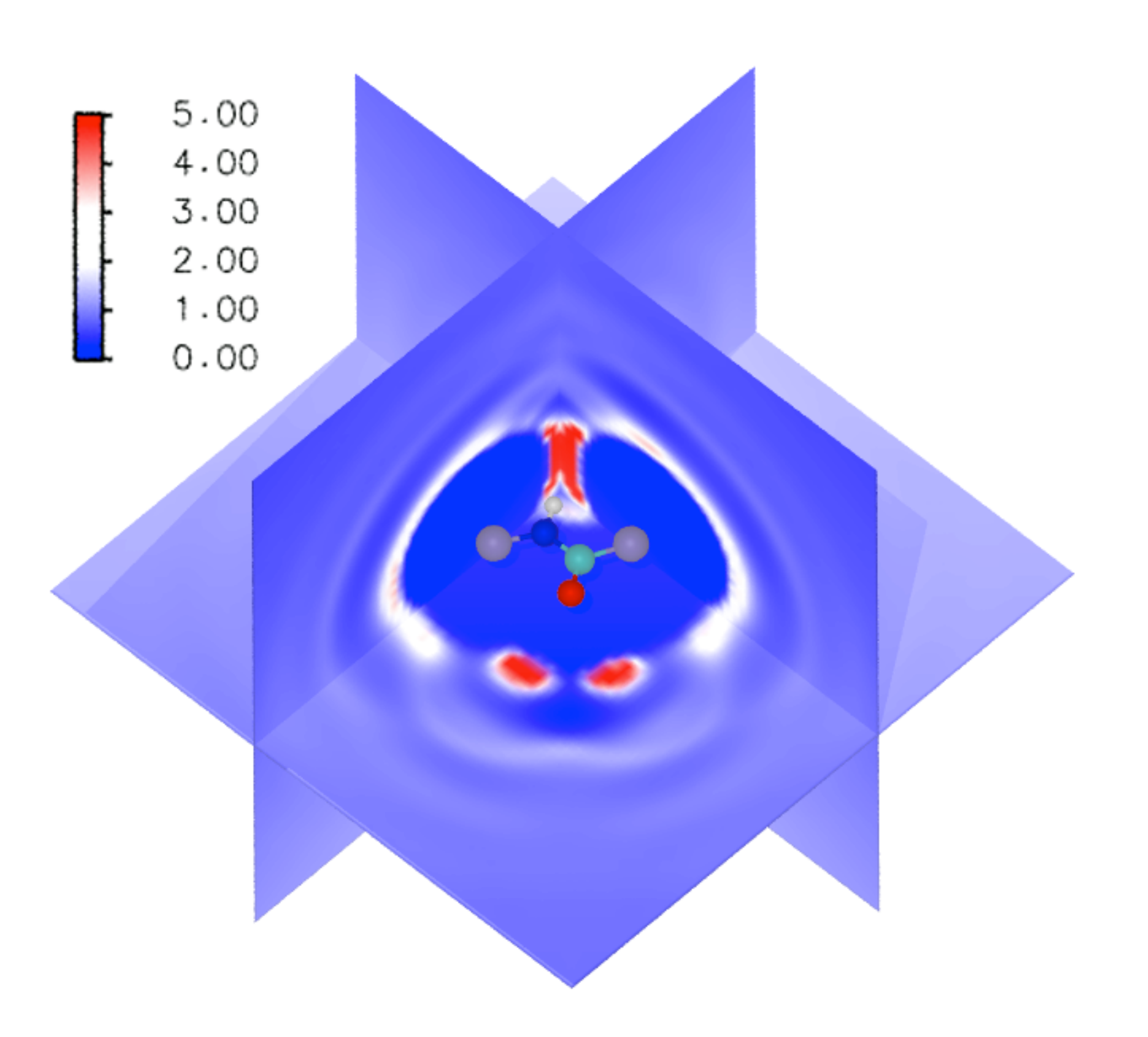}}
\vskip2em
\resizebox{14cm}{!}{\includegraphics{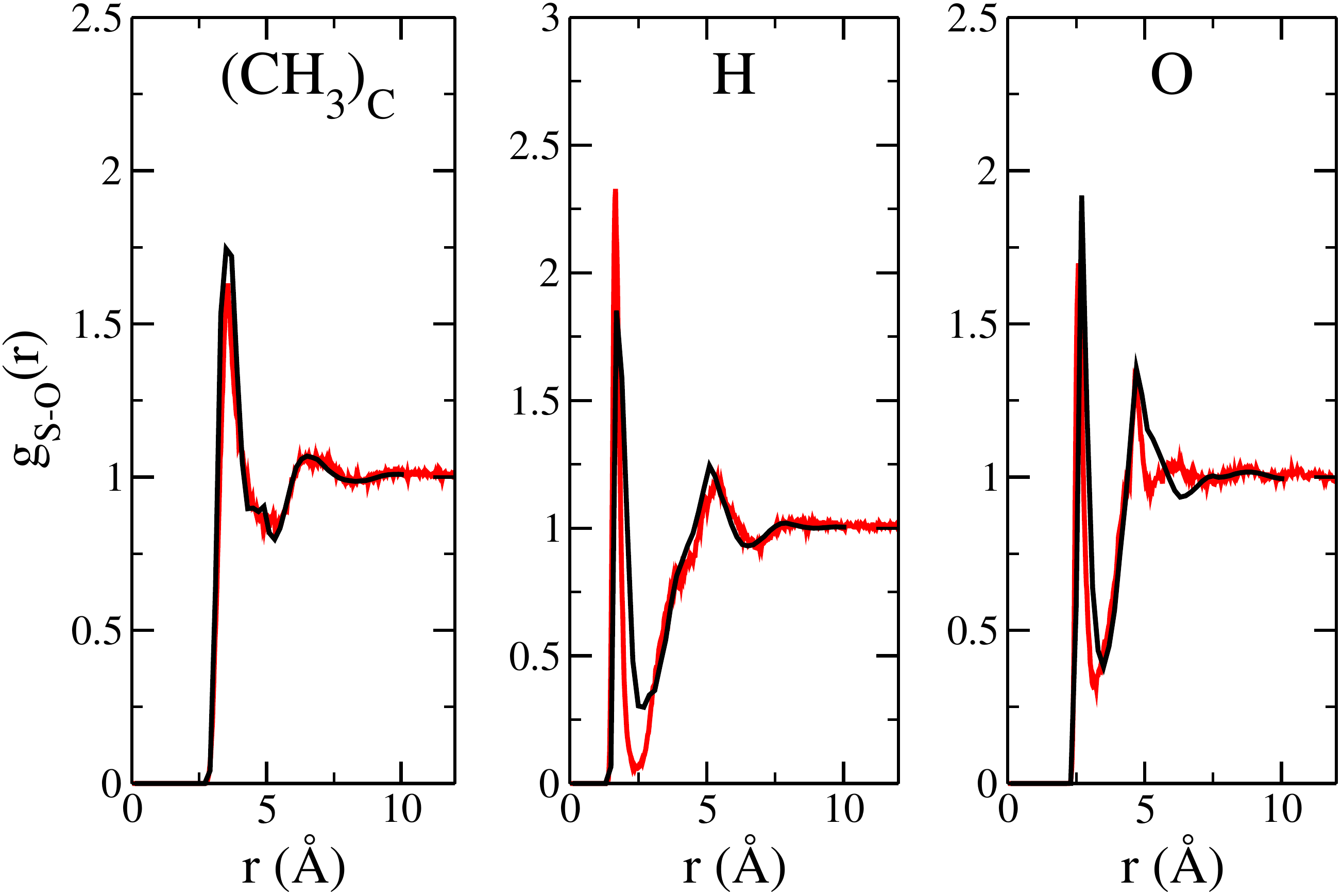}}
\end{center}
\caption{
\label{fig:gr_NMA}
Top: Three-dimensional water density around a  N-methyl-acetamide molecule obtained by MDFT. Bottom: A selection of solute site-oxygen pair distribution functions computed by MD (red lines) or MDFT (black lines). 
}
\end{figure}

\end{document}